\documentclass[onecolumn,showpacs,preprintnumbers,amsmath,amssymb]{revtex4}

\usepackage{graphicx}
\usepackage{bm}

\begin{document}
\title{Gravitational mass in electromagnetic field}
\author{Zihua Weng}
\email{xmuwzh@hotmail.com; xmuwzh@xmu.edu.cn.}

\affiliation{School of Physics and Mechanical \& Electrical
Engineering, Xiamen University, Xiamen 361005, China}

\begin{abstract}
A fraction of energy is theoretically predicted to be captured from
electromagnetic field to form a gravitating mass, when a low-mass
charged particle enters the strong field from a region of no
electromagnetism. In this paper the mass variation has been
calculated for a charged particle on free-fall in the constraint
electromagnetic field. It has been shown that there is an evident
effect to the variation in mass when the low-mass charged particle
is in the strong electromagnetic field.
\end{abstract}


\maketitle

\section{Introduction}

The equality of the inertial and gravitational masses is being
doubted all the time. And the equality of masses remains as puzzling
as ever. However, the existing theories do not explain why the
gravitational mass has to equal the inertial mass, and then do not
offer compelling reason for this empirical fact. The paper attempts
to reason out why there exists the equality of masses in most cases,
even in the electromagnetic field.

Some experiments for the equality of masses have been performed by
L. E\"{o}tv\"{o}s \citep{eotvos}, H. Potter \citep{potter}, R. Dicke
\citep{dicke}, and V. Braginsky \citep{braginsky}, etc. And more
precise experiments have been carried out by I. Shapiro
\citep{shapiro}, K. Nordtvedt \citep{nordtvedt}, and J. Gundlach
\citep{gundlach} etc. Presently, no deviation from this equality has
ever been found, at least to the accuracy $10^{-15}$. But all of
these verifications are solely constrained to be in the range of
weak gravitational strength, and have not been validated in the
strong gravity nor in the electromagnetic field. So this puzzle of
the equality of masses remains unclear and has not satisfied
results.

The paper brings forward a theoretical model to study the variation
of gravitational mass in the electromagnetic field. It carries out
the result that the gravitational mass will be varied in the strong
strength by capture or release the energy density of the
electromagnetic field.

\section{Electromagnetic field}

The electromagnetic field can be described with the quaternion
\citep{hamilton}. In the treatise on electromagnetic field theory,
the quaternion was first used by J. \citet{maxwell} to demonstrate
the electromagnetic field in 1873. And the gravitational field can
be described by the quaternion also, and can be used to work out the
variation of the gravitational mass in the gravitational field
\citep{weng1}.

The gravitational field and electromagnetic field both can be
illustrated by the quaternion, but they are quite different from
each other indeed. It is assumed for simplicity that there exists
one kind of symmetry between the electromagnetic field with the
gravitational field. We add another four-dimensional basis vector to
the ordinary four-dimensional basis vector to include the feature of
the gravitational and electromagnetic fields \citep{weng2}.

Some physical quantities are the functions of coordinates $r_0 , r_1
, r_2$, and $r_3$ in these two kinds of fields, but the basis
vector, $\mathbb{E}_g = (1 , \emph{\textbf{i}}_1 ,
\emph{\textbf{i}}_2 , \emph{\textbf{i}}_3)$, of the gravitational
field differs from the basis vector, $\mathbb{E}_e =
(\emph{\textbf{e}} , \emph{\textbf{j}}_1 , \emph{\textbf{j}}_2 ,
\emph{\textbf{j}}_3)$, of the electromagnetic field. Two kinds of
basis vectors constitute the basis vector $\mathbb{E} = \mathbb{E}_g
+ \mathbb{E}_e$ of the octonion \citep{cayley}, with $\mathbb{E}_e =
\mathbb{E}_g \circ \emph{\textbf{e}}$ . The symbol $\circ$ denotes
the octonion multiplication. And the octonion basis vectors satisfy
the multiplication characteristics in Table \ref{tab:table1}.

In the quaternion space, the radius vector $\mathbb{R}$ is
\begin{eqnarray}
\mathbb{R} = r_0 + r_1 \emph{\textbf{i}}_1 + r_2 \emph{\textbf{i}}_2
+ r_3 \emph{\textbf{i}}_3
\end{eqnarray}
where, $r_0 = ct$~; $c$ is the light's speed; $t$ denotes the time.

\begin{table}[]
\caption{The octonion multiplication table.}
\label{tab:table1}
\centering
\begin{tabular}{ccccccccc}
\hline\hline $ $ & $1$ & $\emph{\textbf{i}}_1$  &
$\emph{\textbf{i}}_2$ & $\emph{\textbf{i}}_3$  & $\emph{\textbf{e}}$
& $\emph{\textbf{j}}_1$
& $\emph{\textbf{j}}_2$  & $\emph{\textbf{j}}_3$  \\
\hline $1$ & $1$ & $\emph{\textbf{i}}_1$  & $\emph{\textbf{i}}_2$ &
$\emph{\textbf{i}}_3$  & $\emph{\textbf{e}}$  &
$\emph{\textbf{j}}_1$
& $\emph{\textbf{j}}_2$  & $\emph{\textbf{j}}_3$  \\
$\emph{\textbf{i}}_1$ & $\emph{\textbf{i}}_1$ & $-1$ &
$\emph{\textbf{i}}_3$  & $-\emph{\textbf{i}}_2$ &
$\emph{\textbf{j}}_1$
& $-\emph{\textbf{e}}$ & $-\emph{\textbf{j}}_3$ & $\emph{\textbf{j}}_2$  \\
$\emph{\textbf{i}}_2$ & $\emph{\textbf{i}}_2$ &
$-\emph{\textbf{i}}_3$ & $-1$ & $\emph{\textbf{i}}_1$  &
$\emph{\textbf{j}}_2$  & $\emph{\textbf{j}}_3$
& $-\emph{\textbf{e}}$ & $-\emph{\textbf{j}}_1$ \\
$\emph{\textbf{i}}_3$ & $\emph{\textbf{i}}_3$ &
$\emph{\textbf{i}}_2$ & $-\emph{\textbf{i}}_1$ & $-1$ &
$\emph{\textbf{j}}_3$  & $-\emph{\textbf{j}}_2$
& $\emph{\textbf{j}}_1$  & $-\emph{\textbf{e}}$ \\
\hline $\emph{\textbf{e}}$ & $\emph{\textbf{e}}$ &
$-\emph{\textbf{j}}_1$ & $-\emph{\textbf{j}}_2$ &
$-\emph{\textbf{j}}_3$ & $-1$ & $\emph{\textbf{i}}_1$
& $\emph{\textbf{i}}_2$  & $\emph{\textbf{i}}_3$  \\
$\emph{\textbf{j}}_1$ & $\emph{\textbf{j}}_1$ & $\emph{\textbf{e}}$
& $-\emph{\textbf{j}}_3$ & $\emph{\textbf{j}}_2$  &
$-\emph{\textbf{i}}_1$
& $-1$ & $-\emph{\textbf{i}}_3$ & $\emph{\textbf{i}}_2$  \\
$\emph{\textbf{j}}_2$ & $\emph{\textbf{j}}_2$ &
$\emph{\textbf{j}}_3$ & $\emph{\textbf{e}}$  &
$-\emph{\textbf{j}}_1$ & $-\emph{\textbf{i}}_2$
& $\emph{\textbf{i}}_3$  & $-1$ & $-\emph{\textbf{i}}_1$ \\
$\emph{\textbf{j}}_3$ & $\emph{\textbf{j}}_3$ &
$-\emph{\textbf{j}}_2$ & $\emph{\textbf{j}}_1$  &
$\emph{\textbf{e}}$  & $-\emph{\textbf{i}}_3$
& $-\emph{\textbf{i}}_2$ & $\emph{\textbf{i}}_1$  & $-1$ \\
\hline
\end{tabular}
\end{table}

The velocity $\mathbb{V}$ is defined as
\begin{eqnarray}
\mathbb{V} = \partial\mathbb{R}/\partial t = c + v_1
\emph{\textbf{i}}_1 + v_2 \emph{\textbf{i}}_2 + v_3
\emph{\textbf{i}}_3
\end{eqnarray}

The gravitational potential is $\mathbb{A}_g = (a_0 , a_1 , a_2 ,
a_3)$, and the electromagnetic potential is $\mathbb{A}_e = (A_0 ,
A_1 , A_2 , A_3)$. The gravitational field and electromagnetic field
constitute the modified gravity field, which potential is
$\mathbb{A}$ .
\begin{eqnarray}
\mathbb{A} = \mathbb{A}_g + k_{eg} \mathbb{A}_e
\end{eqnarray}
where, $k_{eg}$ is the coefficient.

The strength $\mathbb{B}$ consists of the gravitational strength
$\mathbb{B}_g$ and the electromagnetic strength $\mathbb{B}_e$.
\begin{eqnarray}
\mathbb{B} = \lozenge \circ \mathbb{A} = \mathbb{B}_g + k_{eg}
\mathbb{B}_e
\end{eqnarray}
where, the operator $ \lozenge = \partial_0 + \emph{\textbf{i}}_1
\partial_1 + \emph{\textbf{i}}_2 \partial_2 + \emph{\textbf{i}}_3 \partial_3$; and the $\partial_i =
\partial/\partial r_i$, $i = 0, 1, 2, 3 $ .

In the above equation, we choose the following gauge conditions to
simplify succeeding calculation.
\begin{eqnarray}
\partial_0 a_0 + \nabla \cdot \textbf{a} = 0 ~,~~ \partial_0 A_0 + \nabla \cdot \textbf{A} = 0
\end{eqnarray}
where, the operator $\nabla = \emph{\textbf{i}}_1
\partial_1 + \emph{\textbf{i}}_2
\partial_2 + \emph{\textbf{i}}_3 \partial_3~;~ \textbf{a}
= \emph{\textbf{i}}_1 a_1 + \emph{\textbf{i}}_2 a_2 +
\emph{\textbf{i}}_3 a_3~;~ \textbf{A} = \emph{\textbf{i}}_1 A_1 +
\emph{\textbf{i}}_2 A_2 + \emph{\textbf{i}}_3 A_3 $ .

The gravitational strength $\mathbb{B}_g$ includes two components,
$\textbf{g} = ( g_{01} , g_{02} , g_{03} ) $ and $\textbf{b} = (
g_{23} , g_{31} , g_{12} )$ ,
\begin{eqnarray}
\textbf{g}/c = && \emph{\textbf{i}}_1 ( \partial_0 a_1 +
\partial_1 a_0 ) + \emph{\textbf{i}}_2 ( \partial_0 a_2 + \partial_2
a_0 ) + \emph{\textbf{i}}_3 ( \partial_0 a_3 + \partial_3 a_0 )
\nonumber
\\
\textbf{b} = && \emph{\textbf{i}}_1 ( \partial_2 a_3 -
\partial_3 a_2 ) + \emph{\textbf{i}}_2 ( \partial_3 a_1 - \partial_1
a_3 ) + \emph{\textbf{i}}_3 ( \partial_1 a_2 - \partial_2 a_1 )
\nonumber
\end{eqnarray}
meanwhile the electromagnetic strength $\mathbb{B}_e$ involves two
parts, $\textbf{E} = ( B_{01} , B_{02} , B_{03} ) $ and $\textbf{B}
= ( B_{23} , B_{31} , B_{12} )$ .
\begin{eqnarray}
\textbf{E}/c = && \emph{\textbf{j}}_1 ( \partial_0 A_1 + \partial_1
A_0 ) + \emph{\textbf{j}}_2 ( \partial_0 A_2 + \partial_2 A_0 ) +
\emph{\textbf{j}}_3 ( \partial_0 A_3 + \partial_3 A_0 )
\nonumber\\
\textbf{B} = && \emph{\textbf{j}}_1 ( \partial_3 A_2 - \partial_2
A_3 ) + \emph{\textbf{j}}_2 ( \partial_1 A_3 - \partial_3 A_1 ) +
\emph{\textbf{j}}_3 ( \partial_2 A_1 - \partial_1 A_2 ) \nonumber
\end{eqnarray}

The linear momentum $\mathbb{S}_g = m \mathbb{V} $ is the source of
the gravitational field, and the electric current $\mathbb{S}_e = q
\mathbb{V} \circ \emph{\textbf{e}}$ is the source of the
electromagnetic field. The source $\mathbb{S}$ satisfies,
\begin{eqnarray}
\mu \mathbb{S}  = - ( \mathbb{B}/c + \lozenge)^* \circ \mathbb{B} =
\mu_g \mathbb{S}_g + k_{eg} \mu_e \mathbb{S}_e - \mathbb{B}^* \circ
\mathbb{B}/c
\end{eqnarray}
where, $m$ is the mass; $q$ is the electric charge; $\mu$, $\mu_g$,
and $\mu_e$ are the constants; $k_{eg}^2 = \mu_g /\mu_e$ ; $*$
denotes the conjugate of the octonion; $\lozenge^2 = \lozenge^*
\circ \lozenge $ .

The $\mathbb{B}^* \circ \mathbb{B}/(2\mu_g)$ is the energy density,
and includes that of the electromagnetic field.
\begin{eqnarray}
\mathbb{B}^* \circ \mathbb{B}/ \mu_g = \mathbb{B}_g^* \circ
\mathbb{B}_g / \mu_g + \mathbb{B}_e^* \circ \mathbb{B}_e / \mu_e
\end{eqnarray}

The applied force $\mathbb{F}$ is defined from the linear momentum
$\mathbb{P} = \mu \mathbb{S} / \mu_g$ , which is the extension of
the $\mathbb{S}_g$ .
\begin{eqnarray}
\mathbb{F} = c ( \mathbb{B}/c + \lozenge)^* \circ \mathbb{P}
\end{eqnarray}
where, the applied force $\mathbb{F}$ includes the gravity, inertial
force, interacting force between the angular momentum with gravity,
Lorentz force, and interacting force between the magnetic moment
with electromagnetic field, etc.

The above means the field equations of the gravity are modified, but
Maxwell's equations of the electromagnetic field remind unchanged.
Meanwhile the definitions of the applied force and linear momentum
etc. are expanded in the gravitational and electromagnetic fields.

\begin{figure}
\begin{center}
\includegraphics{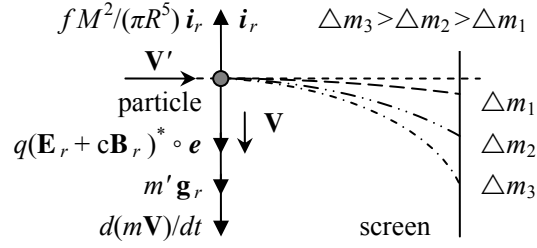}
\caption{In the constraint electromagnetic field which strength
$\textbf{E}_r + c\textbf{B}_r = 0$, the low-mass charged particle
falls freely along the $\emph{\textbf{i}}_r$ and moves uniformly
along the horizontal direction, with $\triangle m$ being $\triangle
m_1$ , $\triangle m_2$ , and $\triangle m_3$ . } \label{fig:figure1}
\end{center}
\end{figure}

In the theoretical model which consists of the electromagnetic and
gravitational fields, the low-mass charged particle is on
equilibrium state, when the total applied force is equal to zero.
The equilibrium equations deduce the free-fall motion of the
low-mass charged particle in Figure \ref{fig:figure1}.

\section{Gravitational mass}

Considering the free-fall motion of a low-mass charged particle
along the earth's radial direction $\emph{\textbf{i}}_r$ in the
electromagnetic field and gravitational field of the earth. We
choose the cylindrical polar coordinate system $(t, r, \theta, z)$
with its origin at the center of the mass of the earth. The z-axis
is directed along the rotation axis of the earth, and the $(r,
\theta)$ plane is constrained to coincide with the equatorial plane
of the earth. The radius vector $\mathbb{R}$ and velocity
$\mathbb{V}$ of the particle are respectively,
\begin{eqnarray}
&& \mathbb{R} = r_0 + \textbf{r} , \textbf{r} = r
\emph{\textbf{i}}_r ; \mathbb{V} = c + \textbf{V} , \textbf{V} = V_r
\emph{\textbf{i}}_r .
\end{eqnarray}
and the strength exerted on the low-mass charged particle is
\begin{eqnarray}
&& \mathbb{B} = \textbf{g}/c + k_{eg}\textbf{E}/c +
k_{eg}\textbf{B}~. ~ \textbf{g} = \textbf{g}_r ~ , ~\textbf{g}_r =
g_r \emph{\textbf{i}}_r ~;
\nonumber\\
&& \textbf{E} = \textbf{E}_r ~, ~\textbf{E}_r = E_r
\emph{\textbf{i}}_r \circ \emph{\textbf{e}} ~; ~ \textbf{B} =
\textbf{B}_r ~, ~\textbf{B}_r = B_r \emph{\textbf{i}}_r \circ
\emph{\textbf{e}} ~. \nonumber
\end{eqnarray}
where, $\emph{\textbf{i}}_r$ is the unitary basis vector.

By Eq.(8), the equilibrium equation of total applied force can be
written as
\begin{eqnarray}
( \textbf{g}_r / c^2 + k_{eg} \textbf{E}_r / c^2  + k_{eg}
\textbf{B}_r / c + \lozenge )^* \circ [ m'c + m \textbf{V} + q (
k_{eg} \mu_e / \mu_g ) ( c + \textbf{V} )  \circ \textbf{\emph{e}} ]
= 0
\end{eqnarray}
where, $m' = m + \triangle m$, $\triangle m = - \mathbb{B}^* \circ
\mathbb{B} / (\mu_g c^2) $; $ \lozenge = \partial/\partial r_0 +
\emph{\textbf{i}}_r \partial/\partial r$ .

In the cylindrical polar coordinates $( t , r , \theta , z )$, the
above can be rearranged according to the gravitational field. In the
basis vector $(1, \emph{\textbf{i}}_r , \emph{\textbf{i}}_\theta ,
\emph{\textbf{i}}_z)$,
\begin{eqnarray}
( \textbf{g}_r / c^2  + \lozenge )^* \circ ( m'c + m \textbf{V}) +
(\textbf{E}_r / c^2 + \textbf{B}_r / c)^* \circ [ q ( c + \textbf{V}
) \circ \textbf{\emph{e}}] = 0
\end{eqnarray}
where, $\textbf{g}_r^* = - \textbf{g}_r ;~ \textbf{E}_r^* = -
\textbf{E}_r ; ~ \textbf{B}_r^* = - \textbf{B}_r $.

It is useful to decompose Eq.(11) along the $\emph{\textbf{i}}_r$ ,
\begin{eqnarray}
\partial(m\textbf{V}) /\partial t - m' \textbf{g}_r - q
(\textbf{E}_r + c \textbf{B}_r ) \circ \emph{\textbf{e}} + f M^2 /
(\pi R^5) \emph{\textbf{i}}_r = 0
\end{eqnarray}
where, $M$ is the mass of the earth, and $f$ is the universal
gravitational constant.

In the paper, if we consider the $\partial (m\textbf{V})/\partial t$
and $m' \textbf{g}_r$ as the force of inertia and gravity
respectively in Eq.(12), then the $m$ and $m'$ will be defined as
the inertial mass and gravitational mass correspondingly. And their
difference is the $\triangle m $, which comes from the energy
density of the gravitational and electromagnetic fields.
\begin{eqnarray}
- \triangle m = f M^2 / ( 4 \pi c^2 R^4 ) +  E_r^2 / ( c^4 \mu_e ) +
B_r^2 / ( c^2 \mu_e )
\end{eqnarray}
where, $R$ is the particle's distance to the earth's center.

When it satisfies the condition, $\textbf{E}_r + c \textbf{B}_r =
0$, the low-mass charged particle is on free-fall. And we can
observe the evident effect of the variation of gravitational mass in
the fairly strong strength of the electromagnetic field, which may
be generated by the laser \citep{kaplan}.
\begin{eqnarray}
d(m\textbf{V}) /d t - m' \textbf{g}_r + f M^2 / (\pi R^5)
\emph{\textbf{i}}_r = 0
\end{eqnarray}
where, $\partial(m\textbf{V})/\partial t$ is written as
$d(m\textbf{V})/dt$, when the time $t$ is the sole variable; $m$,
$q$, and $M$ are the constants; $\textbf{E}_r$ and $\textbf{B}_r$
both are static and uniform.

Similarly, when the particle's charge $q = 0$, the low-mass charged
particle is on free-fall as well. And Eq.(14) can be deduced from
Eq.(11) directly. It asserts the $\triangle m$ of the uncharged
particle (the neutron etc.) will change with the strength of the
electromagnetic field also.

From the above, we find the $\triangle m$ has a limited effect on
the motion of the large-mass particle, because the $\triangle m$ is
quite small. Therefore the equality of masses is believed to be
correct in most cases. However, when there exists a very strong
strength of electromagnetic field, the $\triangle m$ can become very
huge, and then has an impact on the motion of the low-mass charged
particle obviously.

\section{Conclusions}

In the theoretical model, the gravitational mass changes with the
electromagnetic strength also, and has a small deviation from the
inertial mass. This states that it will violate the equality of
masses in the fairly strong strength of the electromagnetic field.

The definition of the gravitational mass equates the gravitational
mass $m'$ with the sum of the inertial mass $m$ and the variation of
gravitational mass $\triangle m$. The $\triangle m$ is related to
the energy density of gravitational field and electromagnetic field.
This inference means that the strong strength of the electromagnetic
field will cause the distinct variation of gravitational mass also.
But this problem has never been discussed before. In other words,
the equality of masses has not been validated in the strong
electromagnetic field.

In contrast to the gravitational field, the super-strong strength of
electromagnetic field may be easier to achieve. If we could obtain
the super-strong electromagnetic field in the lab, more experiments
about the equality of masses would be performed. One international
research group found that the variation of the proton-electron mass
ratio based on lab measurement \citep{reinhold}. It is expected to
find the similar experimental results under different strength of
the electromagnetic field or gravitational field.

It should be noted that the study for the gravitational mass has
examined only one kind of the simple case, of which the low-mass
charged particle is on the free-fall motion in the constraint
electromagnetic field or the gravitational field. Despite its
preliminary character, this study can clearly indicate that the
gravitational mass is always changed with the electromagnetic
strength. For the future studies, the theoretical model will
concentrate on only the suitable predictions about the large
variation of gravitational mass in the fairly strong strength of the
electromagnetic field.

\begin{acknowledgements}
The author is grateful for the financial support from the National
Natural Science Foundation of China under grant number 60677039,
Science \& Technology Department of Fujian Province of China under
grant number 2005HZ1020 and 2006H0092, and Xiamen Science \&
Technology Bureau of China under grant number 3502Z20055011.
\end{acknowledgements}


\begin{thebibliography}{}

\bibitem[Eotvos, Pek\'{a}r, and Fekete(1922)]{eotvos}
     {von E\"{o}tv\"{o}s, L., Pek\'{a}r, D., and Fekete, E.},
     \textit{Ann Phys.} \textbf{68}, 11 (1922).

\bibitem[Potter (1923)]{potter}
     {Potter, H.},
     \textit{Proceedings of the Royal Society of London}. Series A, \textbf{104}, 588 (1923).

\bibitem[Brans and Dicke(1961)]{dicke}
     {Brans, C. and Dicke, R. H.},
     \textit{Phys. Rev.} \textbf{124}, 925 (1961).

\bibitem[Braginsky and V. I. Panov(1972)]{braginsky}
     {Braginsky, V. B. and Panov, V. I.},
     \textit{Sov. Phys. JETP} \textbf{34}, 463 (1972).

\bibitem[Counselman and Shapiro(1968)]{shapiro}
     {Counselman, C. C. and Shapiro, I. I.},
     \textit{Science} \textbf{162}, 352 (1968).

\bibitem[Nordtvedt(1968)]{nordtvedt}
     {Nordtvedt, K.},
     \textit{Phys. Rev.} \textbf{169}, 1014 (1968).

\bibitem[Gundlach and Merkowitz(2000)]{gundlach}
     {Gundlach, J. H. and Merkowitz, S. M.},
     \textit{Phys. Rev. Lett.} \textbf{85}, 2869 (2000).

\bibitem[Hamilton(1843)]{hamilton} {Hamilton, W. R.},
     \emph{Elements of Quaternions},
     {(Longmans, Green \& Co., London, 1866)}.

\bibitem[Maxwell(1864)]{maxwell} {Maxwell, J. C.},
     \emph{A Treatise on Electricity and Magnetism},
      {(Dover Publications, New York, 1954)}.

\bibitem[Weng(2008)]{weng1} {Weng, Z.-H.},
     \emph{arXiv}: 0806.3523.

\bibitem[Weng(2008)]{weng2} {Weng, Z.-H.},
     \emph{arXiv}: 0805.3216.

\bibitem[(1845)]{cayley} {Cayley, A.},
     \emph{The Collected Mathematical Papers}
     {(Johnson Reprint Co., New York, 1963)}.

\bibitem[Kaplan and Shkolnikov(2002)]{kaplan}
     {Kaplan, A. E. and Shkolnikov, P. L.},
     \textit{Phys. Rev. Lett.} \textbf{88}, 074801 (2002).

\bibitem[Reinhold et al.(2006)]{reinhold}
     {Reinhold, E., Buning, R., Hollenstein, U., Ivanchik, A.,
     Petitjean, P., and Ubachs, W.},
     \textit{Phys. Rev. Lett.} \textbf{96}, 151101 (2006).


\end{thebibliography}
\end{document}